# Venus cloud morphology and motions from ground-based images at the time of the Akatsuki orbit insertion

A. Sánchez-Lavega[1], J. Peralta[2], J. M. Gomez-Forrellad[3], R. Hueso[1], S. Pérez-Hoyos[1], I. Mendikoa[1], J. F. Rojas[1], T. Horinouchi[4], Y. J. Lee[2], S. Watanabe[5]

[1] Departamento de Física Aplicada I, Escuela de Ingeniería de Bilbao, Universidad del País Vasco UPV /EHU, Plaza Ingeniero Torres Quevedo, 48013 Bilbao, Spain.

[2] Institute of Space and Astronautical Science (ISAS/JAXA), Sagamihara, Kanagawa, Japan.

[3] Fundació Observatori Esteve Duran, Montseny 46, 08553 Seva, Barcelona, Spain.

[4] Faculty of Environment Earth Science, Hokkaido University, Hokkaido, Japan

[5] Department of Cosmoscience, Hokkaido University, Hokkaido, Japan

*To whom correspondence should be addressed. E-mail: agustin.sanchez@ehu.es

† Partially based on observations obtained at Centro Astronómico Hispano Alemán, Observatorio de Calar Alto MPIA-CSIC, Almería, Spain.

**Abstract**

We report Venus image observations around the two maximum elongations of the planet at June and October 2015. From these images we describe the global atmospheric dynamics and cloud morphology in the planet before the arrival of JAXA's Akatsuki mission on December the 7$^{th}$. The majority of the images were acquired at ultraviolet wavelengths (380-410 nm) using small telescopes. The Venus dayside was also observed with narrow band filters at other wavelengths (890 nm, 725-950 nm, 1.435 μm $CO_2$ band) using the instrument PlanetCam-UPV/EHU at the 2.2m telescope in Calar Alto Observatory. In all cases, the *lucky imaging* methodology was used to improve the spatial resolution of the images over the atmospheric seeing. During the April-June period, the morphology of the upper cloud showed an irregular and chaotic texture with a well developed equatorial dark belt (afternoon hemisphere), whereas during October-December the dynamical regime was dominated by planetary-scale waves (Y-horizontal, C-reversed and ψ-horizontal features) formed by long streaks, and banding suggesting more stable conditions. Measurements of the zonal wind velocity with cloud tracking in the latitude range from 50ºN to 50ºS shows agreement with retrievals from previous works.

**1. Introduction**

Venus clouds show spatial and temporal variability at different levels as well as numerous wave systems. The motions of the clouds and waves can be tracked in detailed images obtained by the different missions and flybys of the planet since 1974 Mariner 10 (Murray et al. 1974; Belton et al. 1976), Pioneer-Venus (Limaye 2007), Galileo (Belton et al. 1991; Peralta et al. 2007), MESSENGER (Solomon et al. 2007) and Venus Express (Sánchez-Lavega et al. 2008; Khatunsev et al. 2013; Hueso et al. 2015). Since December 2015, the Akatsuki orbiter (also named Venus Climate Orbiter) from the Japan Aerospace Exploration Agency (JAXA) continues the exploration of Venus atmosphere (Nakamura et al. 2007; Nakamura et al. 2011; Nakamura et al., 2014). Following a malfunction on the propulsion system in a first orbit insertion attempt, Akatsuki was successfully inserted in an orbit different to that initially planned after maneuvering with the reaction control system (Nakamura et al. 2016). The spacecraft entered Venus orbit (VOI-R1) on December 7 2015 (VOI-R1 VOI from 23:51:29 on December 6 through 00:11:57 on December 7, UTC onboard time, add 8-min 19s for UTC on Earth) and following a trim maneuver at first periapsis, Akatsuki entered inDecember 21 in an orbit with an apoapsis of 360,000-380,000 km, periapsis altitude 1,000-8,000 km and orbital period 10.5 days (Nakamura et al. 2016). Since the Venus Express mission (VEx) ended in November 2014, there are only ground-based observations of the planet during the gap of about a year between the two missions (October 2014 to December 2015).

Ground-based images of Venus clouds are in general scarce in time and they are concentrated in short periods of time due to the intrinsic difficulties to observe the planet from the ground given its angular proximity to the Sun. This restricts the visibility of the illuminated (Venus dayside) or dark (Venus night side) parts of the disk to dates close to its maximum elongation, when the angular distance to the Sun reaches ~ 48° and Venus is half illuminated. The "lucky imaging" technique based on the alignment and stacking of hundreds or thousands of short exposures (Lelièvre et al., 1988; Law et al., 2005), has been successfully employed for high resolution at visible wavelengths in a number of astronomical contexts. It permits high resolution planetary imaging with small telescopes in the range ~0.3-0.5 m(, thus allowing to monitor Venus's clouds in the optical range during practically its whole orbit without the restrictions for Sun distance (Mousis et al. 2014). For this reason, the European Space Agency strongly encouraged the acquisition of Venus images from amateur observers to support early observations by the VEx spacecraft (Barentsen & Koschny, 2008). The evolution of the cloud morphology and their motions in Venus daytime side are typically studied using images taken in the ultraviolet-violet (0.38-0.42 µm, UV) and near infrared (0.9-0.97 µm, NIR) spectral ranges where clouds at two altitude levels can be observed (Belton et al. 1991; Sánchez-Lavega et al. 2008).

In this paper we report the analysis of ground-based observations of the cloud morphology and motions in Venus daytime atmosphere prior to and during the VOI-R1 phase of Akatsuki in 2015 (Eastern elongation of 45.4° in June 7 and Western elongation of 46.4° in October 26, respectively). We used images in the UV from different public databases taken by amateur astronomers. We also present images of the planet taken with the instrument PlanetCam-UPV/EHU (Mendikoa et al. 2016) mounted on the 2.2 m telescope at Calar Alto Observatory (CAHA) that observed simultaneously in the visible (0.38-1 µm) and the SWIR (1-1.7 µm) spectral ranges.

**2. Image selection and measurement method**

We selected Venus images according to their quality (appropriate contrast and spatial resolution) and temporal coverage during the two greatest elongation periods. *($1^{st}$)* pre- Akatsuki VOI (April to June 2015): This was an Eastern elongation with angular distance between the Sun and Venus (*e*), phase angle (α) and Venus diameter ($D$) ranging from $e = 37°$, $\alpha = 55.8°$ and $D = 13.8"$ to $e = 45.4°$, $\alpha = 107.8°$ and $D = 42"$. (*2$^{nd}$*) AkatsukiVOI-R1(October to December 2015): This was a Western elongation with $e$ and $D$ ranging from $e = 46.4°$, $\alpha = 107.8°$ and $D = 33.0"$ to $e = 38°$, $\alpha = 57.5°$ and $D = 14"$. The images were downloaded and are available in the following databases:

(1) Association of Lunar and Planetary Observers (ALPO) – Japan:

http://alpo-j.asahikawa-med.ac.jp/Latest/Venus.htm

(2) Société Astronomique de France (SAF): Commission Observations Planétaires de la SAF – section Venus (Astrosurf):

http://www.astrosurf.com/planetessaf/venus/index.htm

(3) Unione Astrofili Italiani (UAI): Vetrina di osservazioni planetarie

http://pianeti.uai.it/archiviopianeti/

In total we selected and examined about 525 images (340 during the Eastern elongation and 185 during the Western elongation) obtained in the UV-violet band (380-410 nm). The PlanetCam imaging series were obtained in a single observing run on 30 December 2015. Additional support was obtained with images obtained with a Celestron 11" telescope from the Aula EspaZio Gela (Sánchez-Lavega et al., 2014). Observers and times of observations are specified in the captions of the corresponding figures.

Figure 1a shows the Venus dayside reflected spectrum with the main features: UV aerosol absorption (300-450 nm) and $CO_2$ absorption bands at 1.05, 1.2–1.35 and 1.435 µm. Figure 1b-1e shows the aspect of Venus at different wavelengths and Figures 1f-1g are examples of the spatial and temporal resolution of the images used in this study. For a telescope with a diameter of 40 cm the resolution limit imposed by diffraction is σ (380 nm) = 0.25". During greatest elongations, for a Venus diameter ~ 40", the spatial resolution at Venus Equatorial latitudes is ~ 200 km/pix (380 nm). For the 2.2 m telescope at CAHA these numbers are σ (380 nm) = 0.05" and σ (1,430 nm) = 0.17", so at maximum elongations the maximum spatial resolution attainable at Equatorial latitudes is ~ 40 km/pix (380 nm), 100 km/pix (980 nm) and 850 km/pix (1.43 μm).

We used the WinJUPOS software (Hahn, 2016) to navigate the images transforming the original pixel coordinates to longitude and latitude positions over the planet. This was done by orienting the images and fitting the limb and phase of the planet to a synthetic limb and phase based on ephemeris and calculated by the software. The PlanetCam images processed with our pipeline PLAYLIST that evaluates the quality of individual frames, coregisters individual frames and stacks the best frames into the final scientific image (Mendikoa et al. 2016). For the determination of the longitudes on the planet disk two rotation periods are implemented in WinJUPOS: System 1 (S1) is for the planetary surface rotation with period 243.018 days (Archinal et al. 2010) and System 2 (S2) is arbitrarily chosen close to the superrotation period of 4.2 days at the upper cloud layer (equivalent to an angular velocity of –85.7142857°/day) (Crussaire 2004). Cloud features captured in UV images remain quasi-stationary in System 2 and full hemispheric cylindrical maps can be composed in a 4-day period. See e.g. Venus cloud maps from the Mariner 10 flyby based on a 4.0 day rotation period at the Equator (Murray et al. 1974; Belton et al. 1976) and maps from the Galileo flyby based on a rotation period of 4.4 days (Kouyama et al. 2012). In this case, we used the WinJUPOS software to generate mosaics maps in System 2 representative of the global cloud morphology over the whole planet.

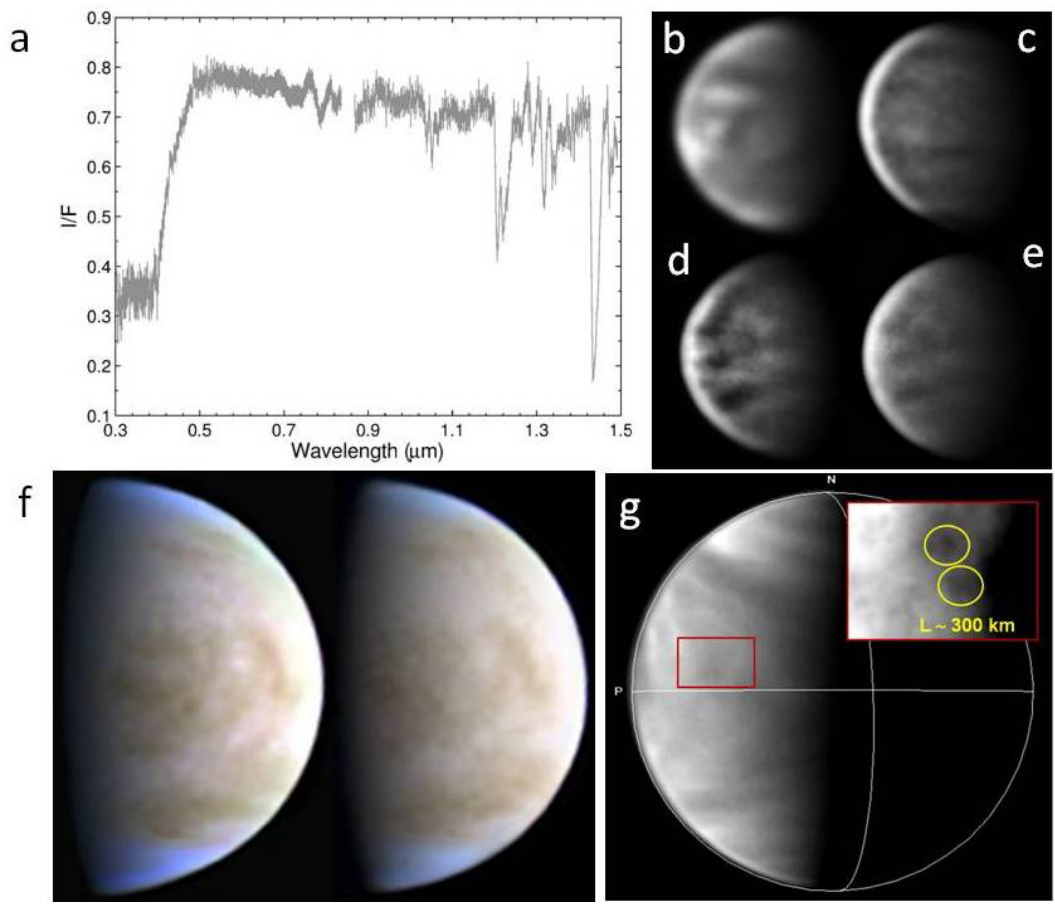

**Figure 1.** Venus spectra and images. (a) Venus dayside spectra (0.3-1.5μm) from Messenger MASCS instrument (VIRS spectrograph) obtained during the flyby on June 5, 2007 (Perez-Hoyos et al., 2013). (b, c, d, e) PlanetCam images obtained with the 2.2 m Calar Alto Obs. telescope on December 30, 2015 at wavelengths: (b) 380 nm (06:35 UT); (c) 1.435 μm (06:54 UT); (d) 890 nm (06:56 UT); (e) 725-950 nm (06:50 UT); (f) Color composite (dominated by UV absorption) images separated by 2 hr (15:14 UT on the left and 17:10 UT on the right) taken by V. Alekssev (Lipetsk, Russia) with a telescope of 40 cm on May 20, 2015 with a Venus diameter D = 19.7", elongation e = 44.5°; (g) UV image taken in November 14, 2015 at 22:50 UT by T. Olivetti (Bangkok,

Thailand) with a telescope of 41 cm diameter with a Venus diameter D = 20" and elongation e = 45.4°. The inset show resolved cloud figures with sizes of ~ 300 km. North is up and West to the left (atmospheric rotation from right to left).

## 3. UV Cloud morphology

### 3.1 Eastern elongation (April-June, 2015)

We show in Figure 2 a representative group of images corresponding to the East elongation of Venus during April – June 2015 (see also Figure 1f). During this period the cloud morphology was dominated by the presence of a marked *dark equatorial band* (Figure 2b-c) similar to those observed during the Mariner 10 flyby (Belton et al. 1976; Schubert 1983), Pioneer-Venus (Rossow et al. 1980) and VEx (Titov et al., 2012) and by *arc-shaped features* (Figure 2d-f) resembling the *bow-shape* features reported by Belton et al. (1976) and Rossow et al. (1980). The equatorial latitudes showed a mottled aspect that suggests the presence of *cells* similar to those seen by previous space missions (Rossow et al. 1980; Belton et al. 1991). They typically have sizes of 200-300 km, i.e. at the resolution limit of the best available images (Figure 2 b-c, e-f). During June 2015 there were patterns of patchy dark albedo bands crossing North-South the Equatorial area accompanied by quasi-parallel segments (Figure 2d-e-f). These dark features have horizontal lengths between ~ 1,500–3,000 km and could represent wave formation whereas the patchy brighter spots (Figure 2f) suggest local instability occurring equatorward of mid-latitudes. A local time dependence of this activity can be involved since during Eastern elongations we observed the afternoon side of Venus (LT

= 12 – 24 hr) when a similar behavior in the clouds as been previously reported by (Titov et al., 2012).

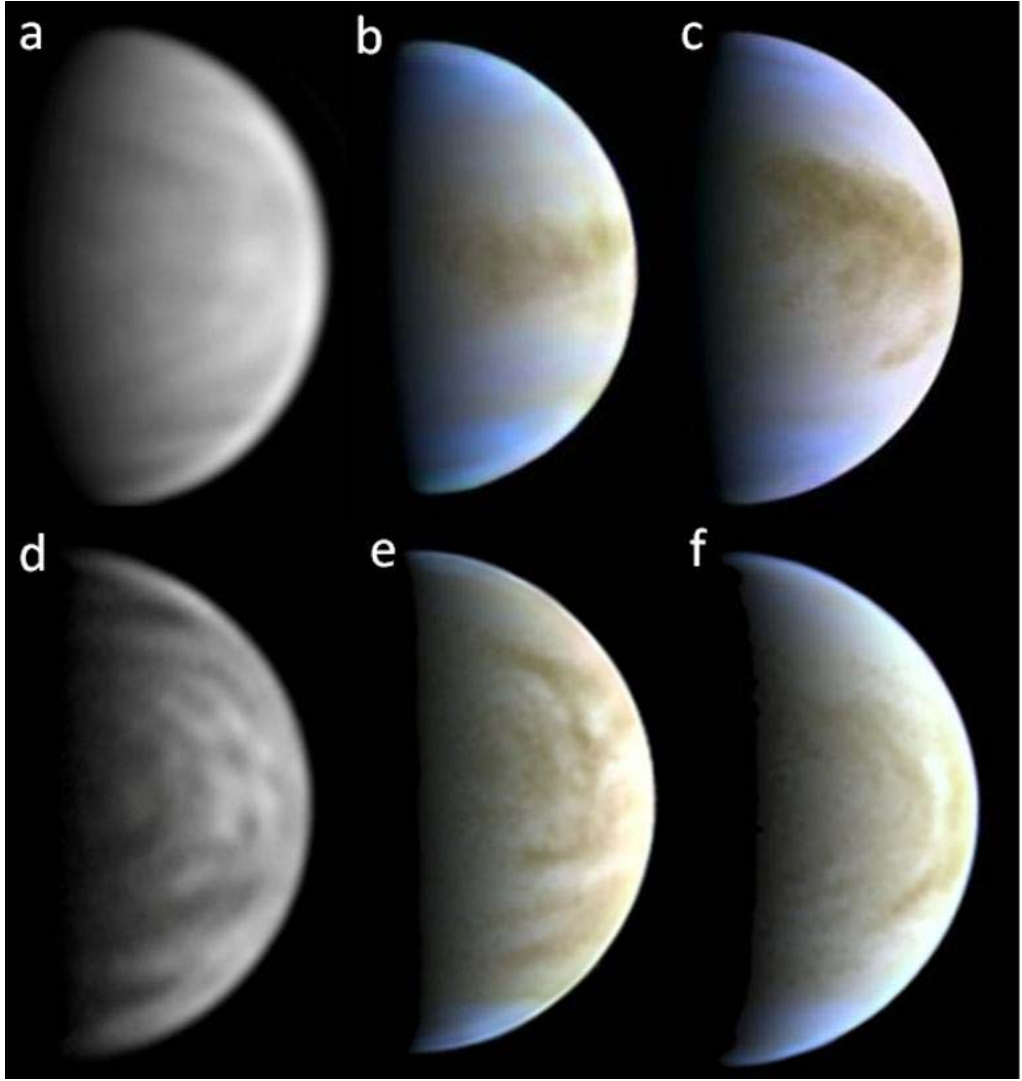

**Figure 2.** Venus cloud morphology in UV during the Eastern elongation. Pre Akatsuki's VOI-R1 (April-June 2015): (a) April 19 (12:06 UT); (b) May 19 (16:04 UT); (c)May 22 (16:34 UT),(d) June 2 (18:42 UT), (e) June 7 (15:18 UT), (f) June 13(15:47 UT)**.** Color compositions (b,c,e,f) are from Red-Green-UV wavelengths with albedo

contrast dominated by UV absorption. Images from: T. Olivetti (a), V. Alekseev (b,c,e,f), D. Gasparri (d).

### 3.2 Western elongation (October - December, 2015)

Figure 3 shows the cloud morphology from October to mid November during the Western Venus elongation. The UV cloud morphology this time (Figure 4) was different to that in April-June showing the characteristic UV-patterns as observed for example during the Pioneer-Venus period described by Rossow et al. (1980). Among these features is the *V-vertex* of the *Y-horizontal wave* that extends in a *dark equatorial band*. This structure showed prominently in images taken on 24 October (Figure 3a) and 27-28 October (Figure 3b) with the equatorial band between latitudes 18ºN and 8ºS extending along the Equator more than 14,000 km in longitude. Figure 3b shows the *V-vertex* of the two arms of the *Y-horizontal* structure spanning from 40ºN to 43ºS. A comparison of the *V-vertex* on 24 and 28 October (Figures 3a and 3b) shows important changes in its morphology after one full rotation (4 days). Only some of the major albedo markings survive and can be used (in addition to the vertex itself) to measure their velocity.

Figure 3c shows a full hemispheric map compiled from images obtained from 13 to 16 November where the *Y-horizontal wave* (System 2 longitudes ~ 180º-360º) and its accompanying *C-reversed wave* (System 2 longitudes ~ 0º-180º) stand prominently, spanning the full range of longitude in the planet. This is a manifestation of the planetary-scale wave pattern with a zonal wavenumber 2 and was particularly conspicuous during that period. The *Y-horizontal* wave and similar features (nicknamed

*C-reversed* and *Psi-horizontal*) have been observed regularly with ground-based telescopes (e.g. Dollfus 1975) and from orbit by different space missions (Murray et al. 1974; Belton et al. 1976; Schubert 1983; Rossow et al. 1980; Peralta et al. 2007, Titov et al. 2012). Particularly interesting is the morphology of the *C-reversed wave* that was meridionally stretched along the longitude sector ~ 70º - 150º (Figure 3c). This mode showed a shorter vertex in longitude than that of the *Y-horizontal wave*, and in the northern side the arm reached high latitudes ~ 55ºN - 60ºN. This meridional stretching is consistent with the evolution of the wave as it interacts with the zonal flow as predicted by the cyclostrophic Kelvin wave model of Peralta et al. (2015). The northern arm exhibited a triple band (longitude sector 70º - 150º) called *spiral dark streaks and polar rings* in Murray et al. (1974) with a North-South separation between them of 900 km. This multiple banding is suggestive of the manifestation of an additional wave, possibly coupled to the Kelvin wave, that requires further analysis.

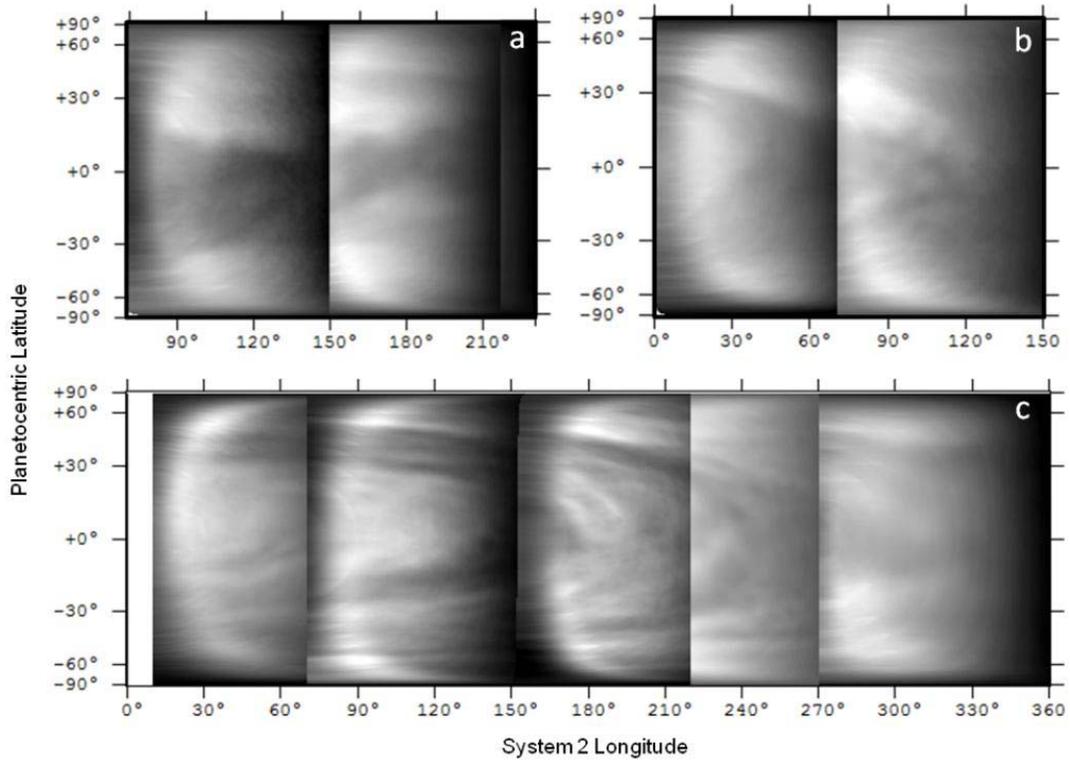

Figure 3. Venus cloud morphology in UV during the Western elongation. Pre Akatsuki's VOI-R1 (October-November 2015) cylindrical projected maps for: (a) 24 October; (b) 27-28 October; (c) 13-16 November. Maps are composed from the following images from left to right: (a) October 24 (05:24 UT, D. Kananovich; 23:13.7 UT, T. Olivetti); (b) October 27 (13:49 UT, P. Maxon), October 28 (5:45 UT, D. Kananovich); (c) November 13 (06:18 UT, R. Braga), Nov. 13 (22:45 T. Olivetti), Nov. 14 (22:50 UT, T. Olivetti), Nov. 15 (06:36 UT, L. Pologni), Nov. 16 (05:40 UT, R. Sedrani).

Figure 4 shows the cloud morphology during the Akatsuki's VOI-R1 epoch from the end of November to mid-December 2015. For reference, Akatsuki first images with different instruments were taken on December 7, 2015 from 4:51:56 to 5:26:02 (UT) (Nakamura et al., 2016). The quality of the available images was not as good as before because of the smaller elongation and size of the planet. However, some interesting features where observed. To facilitate comparison with Akatsuki images we include the System 1 longitude ranges covered by each map forming the map mosaic. First, the *Y-horizontal wave* was again present during this period with the *V-arms* well developed (System 2 longitude range 320º-80º in Figure 4a; S2 longitude range 120º-290º in Figure 4c). Second, the *C-reversed* or perhaps a *ψ-horizontal wave* (in fact a Y-mode with a central equatorial belt) was present in 9-12 December in the S2 longitude range 0º-120º (Figure 4c) accompanying the *Y-horizontal wave*. In addition, a peculiar wavy structure was captured on December 4 (three days before Akatsuki orbit insertion) as shown in Figure 4b. It consisted of a four tilted streak pattern converging toward the Equator at longitude 150º between latitudes 30ºN and 30ºS, then diverging from it apparently as a consequence of the displacement of the bright polar areas toward Equator. This feature can also be partially seen in the UV image obtained by the UVI camera onboard Akatsuki in December 7 (Fig. 6b in Nakamura et al. 2016), and is suggestive of the interaction between the polar vortices edges and the planetary scale waves but seems difficult to assign it to the *Y-*, *ψ-* or *C-* modes.

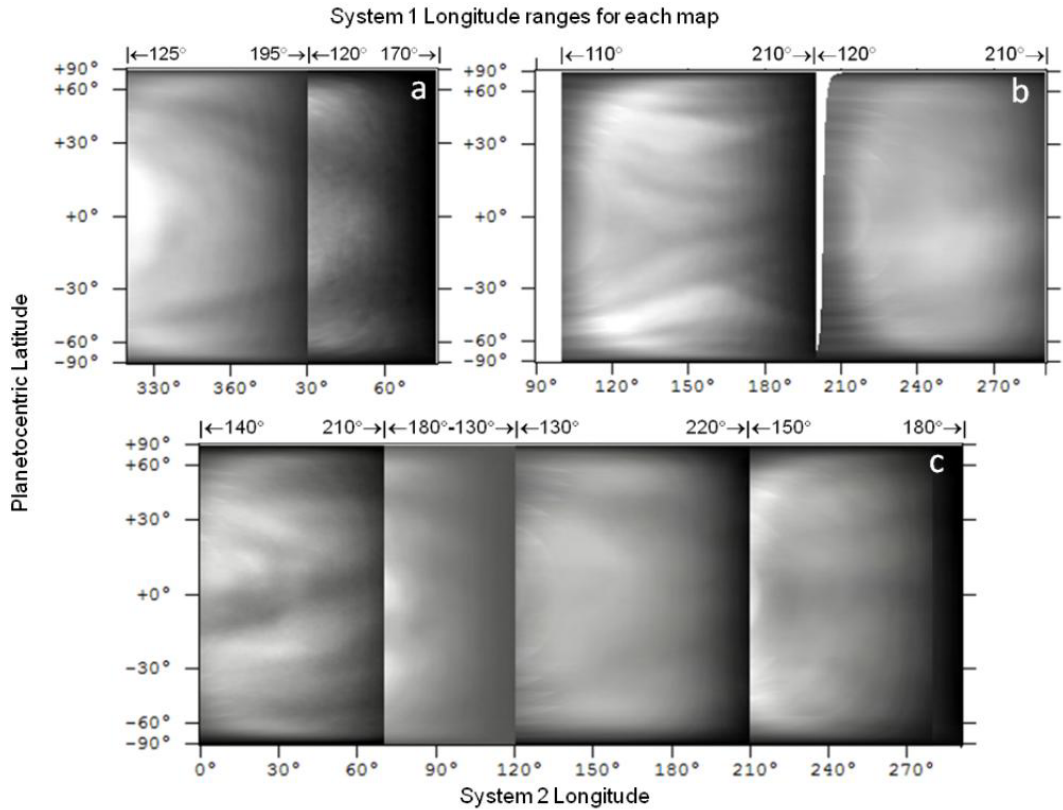

**Figure 4.** Venus cloud morphology in UV at the epoch of Akatuski VOI-R1. Cylindrical projected maps for: (a) 28-29 November, (b) 4-6 December; (c) 9-12 December. To facilitate the comparison with Akatsuki images, System 1 longitude range is indicated in the upper part. Maps are composed from the following images, from left to right: (a) November 28 (23:05 UT, T. Olivetti), November 29 (13:45 UT, P. Maxon); (b) December 4 (23:15 UT, T. Olivetti), December 6 (06:13.5 UT, R. Sedrani); (c) December 11 (23:29, T. Olivetti), December 12 (9:08 UT, L. Pologni), December 9 (07:10 UT, R. Sedrani), December 10 (05:57 UT, R. Sedrani).

On 30$^{th}$ December 2015, following the Akatsuki orbital insertion, we obtained a multi-wavelength set of Venus dayside images with PlanetCam-UPV/EHU (Mendikoa et al., 2016) at the 2.2 m telescope in Calar Alto Observatory (Figure 1b-e). The spatial resolution was low due to the low maximum elevation of the planet at sunrise (26º) but they clearly show the different cloud morphologies depending on the wavelengths (as a consequence of the scattering and absorption properties of the upper cloud). At 380 nm we sense the morphology at altitudes 65-70 km as traced by the mixture of the unknown UV aerosol absorber and the clouds (Figure 1b). In the narrow band at 890 nm and in the broadband 725-950 nm filters (Figure 1 d-e) we sense the cloud morphology at the base of the upper cloud (altitude ~ 60 km; Belton et al. 1991; Sánchez-Lavega et al. 2008). In the 1.435 µm $CO_2$ absorption band (Figure 1c) the cloud morphology differs from that of the previous cases. The vertically integrated extinction within this band indicates that we also sense at this wavelength the upper cloud at 65-70 km but the difference with the UV is that there is very little aerosol absorption and the brightness contrast in the clouds originates fro small differences in altitude and in the scattering behavior of the cloud particles. We measured a relative brightness contrast between features ~ 10% in the UV (380-410 nm) decreasing to about 5% in the NIR (750-950 nm) and SWIR (1.435 µm $CO_2$ band ranges).

## 4. Zonal wind velocity profile

The UV-violet images obtained along the two elongation periods in 2015 were used for cloud tracking and zonal velocity measurements. A standard method was used to retrieve the velocity of individual features as described in Sánchez-Lavega et al. (2008). When high resolution images and adequate sampling are available we use image pairs or triplets separated by a temporal interval $\Delta t = 2 - 6$ hr. This allows to measure wind speeds with errors in the range ~ 20-40 m s$^{-1}$ for tracers with sizes ~200-400 km (see section 2). The sampling involves images obtained from more than one observer, sufficiently separated in Earth longitude, to extend the typical observing window of ~ 2 hr for a single observer. The major source of uncertainty of the method comes from the correct identification of the target and cursor pointing on due to their size and brightness differences them.

A second method we used was the cloud tracking of the major features (sizes ~1,000 km) based on image pairs separated by Δt ~ 4 days. In this particular case, the wind speeds can be measured with smaller errors of ~ 10-15 m s$^{-1}$. However, only major features maintain coherencelong enough to guarantee identification after 4 days. This is the case of features pertaining to the *planetary scale waves* described above. In total, we were able to track 94 cloud features between latitudes ~ 50°N and 50ºS, as shown in Figure 5. For control and checking, we include some measurements (33 wind data points) obtained from additional measurements we performed on ground-based images taken between 2007 and 2012 and retrieved from the same database.

Our data provide the first wind measurements in the UV (upper cloud) from cloud tracking in the Northern hemisphere of Venus since the Galileo flyby in 1990 (Belton et al. 1991; Peralta el al. 2007) because Venus Express wind coverage of the Northern hemisphere was poor. Other long-term ground-based wind measurements encompassing this period for the lower clouds (48-55 km), observed in Venus nighttime (wavelengths 1.7 and 2.3 μm) in both hemispheres, have also been reported (see e.g. Young et al., 2010a,b) Our velocity measurements agree at lower latitudes with results from older missions (Schubert 1983), Doppler wind measurements (Machado et al. 2012, 2014), and the mean profiles from data obtained over a decade by the VMC and VIRTIS instruments onboard VEx (Sánchez-Lavega et al. 2008; Khatuntsev et al. 2013;

Hueso et al. 2015), as shown in Figure 5. The higher discrepancies found for higher latitudes might be attributable to the sporadic changes in the poleward decay of the winds also found in some orbits of VEx mission (see Fig. 4 in Hueso et al. 2015 and Fig. 15 in Khatuntsev et al. 2013) and in the difficulty in target identification at these latitudes.

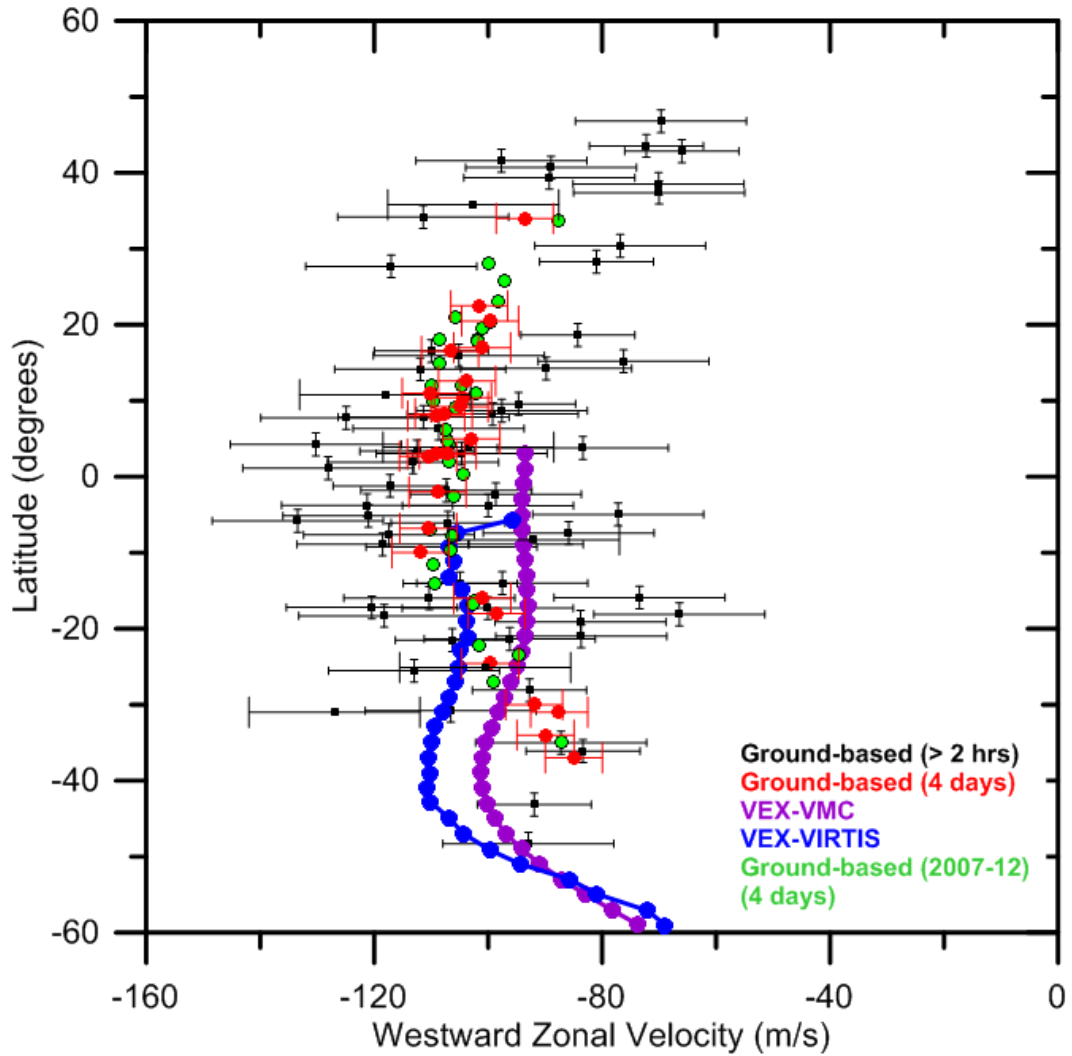

**Figure 5.** Venus zonal wind velocity profile from cloud tracking in UV-violet images. Data are from periods April-May and October-December 2015: Short-term tracking (2-6 hr; black dots), long-term tracking (4 days; red dots). Additional data for 2007-2012 is included (green dots). For comparison we show the average wind profiles obtained with VEx instruments VMC (violet; Khatuntsev et al., 2013) and VIRTIS (blue, Hueso et al., 2015) and representative of the 2006-2012 period.

5. Conclusions

A large observing effort by the amateur astronomer community contributing images to different observational databases has allowed a surveyof the cloud morphology, motions and dynamical state of the upper clouds of Venus as detected in the UV-violet spectral range along 2015. Our main conclusion is that the observed dynamical state of Venus upper cloud differed in April-June from October-December when the Akatsuki VOI took place. During the VOI, the dynamical regime was dominated by the development of planetary-scale waves typical of Venus atmosphere at the cloud top level (*Y-horizontal*, *C-reversed* and *ψ-horizontal waves*) formed by long tilted streaks and banding with complex interactions between them. Part of this difference between both observing periods can be attributed to Local Time observing conditions: afternoon hemisphere and more unstable upper cloud layer (Eastern elongation) and morning with more stable conditions (Western elongation). The zonal wind velocity retrieved from cloud tracking in the latitude range from 50ºN to 50ºS agrees in general with previous wind retrievals although deviations are found. These deviations might be attributable to motions of global-scale waves apparent on the cloud tops and the solar tide (Hueso et al., 2015).

Ongoing ground-based support to Akatsuki spacecraft includes an international observing campaign around the next Eastern elongation in January 2017 with further coordinated observations along the mission development (see the Akatsuki project coordination with ground-based observations: https://akatsuki.matsue-ct.jp/).

**Acknowledgements**

We acknowledge Christophe Pellier and Marc Delcroix (Société Astronomique de France, SAF), and Dr. Paolo Tanga and Raffaello Braga (Unione Astrofili Italiani) and Dr. Dima Titov for their support in the localization of amateur images and observers. This work was supported by the Spanish MICIIN projects AYA2015-65041 with FEDER support, Grupos Gobierno Vasco IT-765-13, and UFI11/55 from UPV/EHU. J.P. acknowledges JAXA's International Top Young Fellowship.

21